\documentclass[%
reprint,amsmath,amssymb,aps,prl,superscriptaddress,longbibliography,
floatfix,
nobalancelastpage
]{revtex4-2}

\usepackage{graphicx}
\usepackage{dcolumn}
\usepackage{bm}
\usepackage{epstopdf}
\usepackage{float}
\usepackage{braket}
\usepackage{dsfont}
\usepackage{amsmath}
\usepackage{siunitx} 
\usepackage[italicdiff]{physics}
\usepackage[T1]{fontenc}
\usepackage[dvipsnames]{xcolor}
\usepackage{color,soul}
\usepackage{lipsum}
\usepackage[colorlinks=true, allcolors=blue]{hyperref}

\usepackage{wrapfig}
\graphicspath{{figures}}

\begin{document}

\title{Phase locking nuclear spins in silicon with spin-orbit coupling}

\author{Habitamu Y. Walelign}
\thanks{These authors contributed equally.}

\author{Manas Ranjan Sahu}
\thanks{These authors contributed equally.}

\affiliation{Department of Physics and Astronomy, University of Rochester, Rochester, NY 14627, USA}
\affiliation{University of Rochester Center for Coherence and Quantum Science, Rochester, NY 14627, USA}

\author{John M. Nichol}
\email{john.nichol@rochester.edu}
\affiliation{Department of Physics and Astronomy, University of Rochester, Rochester, NY 14627, USA}
\affiliation{University of Rochester Center for Coherence and Quantum Science, Rochester, NY 14627, USA}

\begin{abstract}
Because they have such long coherence times, nuclear spins have extraordinary potential for use in quantum information processing devices. 
However, coherent nuclear spin control generally requires external phase references, such as microwave control fields. Here, we phase-lock a $^{29}$Si nuclear spin ensemble in a silicon quantum dot using only the internal electronic spin-orbit coupling as a phase reference. When driven with the quantum-dot electrons, the nuclear spins align themselves to a phase determined by the electronic spin-orbit coupling and the timing of the drive protocol. 
This enables us to measure the coherent precession and inhomogeneous dephasing of the nuclear spins. We  corroborate our results with detailed numerical simulations of the many-body electron nuclear system. Our work opens new routes for coherently controlling solid-state nuclear spin ensembles.

\end{abstract}

\pacs{}

\maketitle

\section{Introduction}
Spin ensembles hold significant promise as valuable elements of quantum sensors~\cite{degen2017quantum}, computers~\cite{wu2010storage,schuster2010high,kubo2010strong,wesenberg2009quantum}, and networks~\cite{azuma2023quantum,appel2025many}. The collective evolution of spins in the presence of external fields enables more precise magnetic field sensing than individual spins; the collective coupling of spins to other quantum elements, such as qubits or resonators often features enhanced coupling; and collective excitations in spin ensembles can be used to store information from other quantum devices. Nuclear spin ensembles, in particular, have excellent potential for these applications because they interact weakly with their environments and have long coherence times~\cite{zhong2015optically,wang2025nuclear,saeedi2013room}. 

However, the weak interaction of a nuclear spin with its environment also creates challenges for spin control with external fields. Beyond nuclear magnetic resonance techniques, the electron-nuclear hyperfine coupling offers a convenient set of control methods. In these techniques, {\it external} fields applied to electron spins, such as electric, magnetic, or optical fields, control the electron spin to manipulate the nuclei~\cite{Sanada2005Gate,Stefan2014Electrically,Gammon2001Electron,Xu2009Optically,gangloff2019quantum,Gillard2022Harnessing}. These coherent spin control approaches generally rely on phase references associated with external electronic control fields.

In this work, we study the effect of spin-orbit coupling, an effective {\it internal} electronic control field, on electron-nuclear interactions. In particular, we show that the presence of electronic spin orbit coupling allows us to both polarize and phase-lock a nuclear spin ensemble in a silicon quantum dot, using central electronic spins. Our process establishes an internal control method for nuclear spins that does not require microwaves or external phase reference.  Instead, the key ingredient of our phase-locking procedure is the electronic spin-orbit coupling~\cite{stepanenko2012singlet,Rudner2010Phase-sensitive}  associated with the semiconductor heterostructure. This spin-orbit coupling acts as a phase reference to which the nuclear spins lock during dynamic nuclear polarization. We tune the nuclear spin phase by adjusting the parameters of our protocol, and measure coherent oscillations and inhomogeneous broadening of the nuclear spin ensemble. Detailed numerical simulations of the many-body electron-nuclear system support our experimental observations.

Our work confirms long-standing predictions~\cite{Brataas2011Nuclear,Brataas2012Dynamical} about the behavior of nuclear spin-ensembles coupled to central spins, and highlight the suprising effects electron spin orbit coupling can have on nuclear spin ensembles. Here we demonstrate nuclear-spin phase locking, but other predictions suggest that topological phase transitions~\cite{Rudner2009Topological,Rudner2010Phase} and even enhanced nuclear spin polarization~\cite{Rudner2010Phase-sensitive} may be possible when both spin-orbit and hyperfine interactions occur. The results presented here lay the foundation for exploring these and other interesting properties of the nuclear spin ensembles.  

\section{Setup and Hamiltonian}
\begin{figure*}[t]
{\includegraphics[width=1\linewidth]{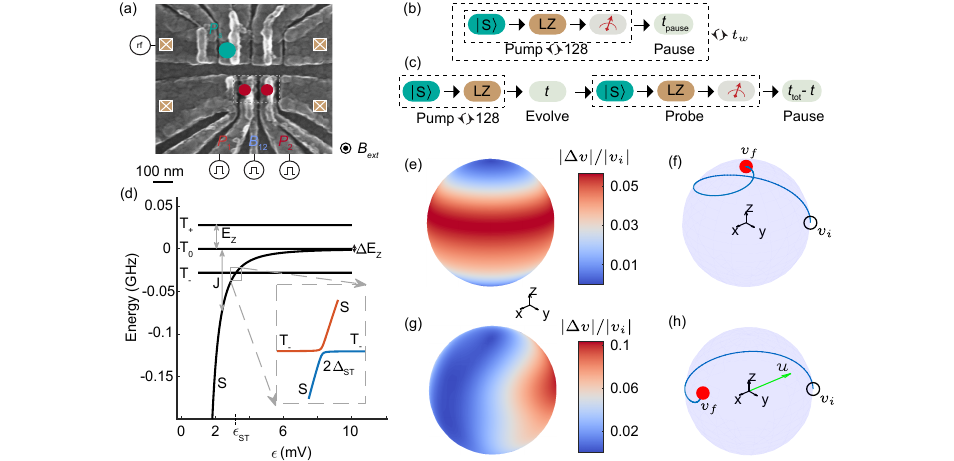}}
\caption{
\textbf{Experimental setup.}
(\textbf{a}) Scanning electron micrograph of a device nominally identical to the one used in this work. The plunger gates $P_{1(2)}$ define the ST qubit, and the gate $P_s$ defines the sensor dot. (\textbf{b}) Pulse sequences used for probing the screened dark states and (\textbf{c}) for the phase control experiment. (\textbf{d}) Relevant energy levels for the experiment. Zeeman energy, $E_z$, separates the two polarized triplet states, $T_\pm$, from the unpolarized triplet state $T_0$. $\Delta E_z$ denotes the longitudinal Zeeman field gradient between the dots while $J$ is the effective exchange coupling that separates the states $S$ and $T_0$. (\textbf{e}) Net change in the nuclear polarization vector after one LZ sweep and pause, as a function of orientation. In the absence of spin-orbit coupling, the blue stable regions point along the $z$-axis and feature vanishing transverse components. (\textbf{f}) Trajectory of the polarization vector under repeated LZ sweep and pause cycles, starting from an initial transverse (open circle, $v_i$) and ending at a final longitudinal state (red dot, $v_f$). Repeated LZ sweeps rotate $v$ toward the $z$-direction, creating longitudinal polarization at the expense of the transverse components of $v$.  (\textbf{g}) In the presence of non-zero spin orbit coupling, additional stable regions are created opposite to the direction of the $u$. (Here $u$ points along the $-x$ direction.) (\textbf{h}) Repeated sweep-and-pause cycles cause $v$ to orient opposite to $u$, when $t_w=t_L$.
}
\label{fig:Stability}
\end{figure*}
We perform our experiments using a gate-defined double dot fabricated using the overlapping-gate architecture~\cite{Angus2007Gate,Zajac2016Scalable} on a natural-abundance Si/SiGe heterostructure with a Si quantum well 50-nm below the surface. The nuclear spins we manipulate in this work are $^{29}$Si nuclear spins in the Si quantum well. We cool the device to below 100~mK using a dilution refrigerator and tune the gate voltages to form the double quantum dot. An adjacent quantum dot is used as a charge sensor optimized for radio-frequency reflectometry based charge sensing~\cite{Connors2020Rapid}[Fig.~\ref{fig:Stability}(a)]. 

We configure the double dot as a four-electron singlet-triplet (ST) qubit~\cite{harvey2017coherent,connors2022charge} with basis states $\ket{S}=\frac{1}{\sqrt{2}}(\ket{\uparrow\downarrow}-\ket{\downarrow\uparrow})$ and $\ket{T_-}=\ket{\downarrow \downarrow}$[Fig.~\ref{fig:Stability} (d)]. Treating the nuclei as classical spins, and up to a constant energy shift, the Hamiltonian for the ST qubit in the $\{\ket{T_-},\ket{S}\}$ basis is
\begin{equation}
    H=\frac{1}{2}\begin{pmatrix}
    \epsilon+v_z & v_++u\\
    v_-+u & -\epsilon-v_z
\end{pmatrix}
\label{eq:matrix}
\end{equation}
Here $v_{\pm}\equiv v_x\pm iv_y$ is the {\it difference} in transverse nuclear polarization between dots and $v_z$ is the {\it total} longitudinal nuclear spin polarization magnitude~\cite{Taylor2007Relaxation}. (See the Supplementary Material for further details on the system Hamiltonian.) The unique nature of the hyperfine matrix elements in this system results from the fact that the relevant electron states are two-spin states that differ both in their total and longitudinal spin angular momentum.  $u$ results from electronic spin-orbit coupling and is tunable through the strength and direction of the external magnetic field~\cite{Tanttu2019Controlling,Ferdous2018Valley}. 
Finally,  $\epsilon$ defines a gate-voltage-controlled detuning, which is related to the voltage-controlled exchange-coupling between electrons. In the following, we define the total coupling matrix element as 
$\Delta_{ST}\equiv v_{\pm}+u$.

To intuitively understand how this system will evolve, let us rewrite Eq.~\ref{eq:matrix} as
\begin{equation}
        H=\vec{S}\cdot(\vec{v}+\vec{B}) \label{eq:ham}.
\end{equation}
Here $\vec{S}=\frac{1}{2}[\sigma_x, \sigma_y, \sigma_z]^{T}$ is a vector of Pauli operators representing the ST qubit, $\vec{v}=[v_x, v_y, v_z]^{T}$ is a vector that represents the net nuclear spin polarization, and the effective field $\vec{B}=[u,0,\epsilon]^T$.  We emphasize that while we focus on ST qubits in this work, the physics we explore is general to any system featuring a Hamiltonian similar to Eq.~\eqref{eq:ham}, including single electrons in self-assembled and gate-defined quantum dots~\cite{Latta2011Hyperfine,Zhao2019Single}, and nitrogen vacancy centers~\cite{Hanson2008Coherent}, for example. While the physical meaning of $\vec{S}$, $\vec{v}$, and $\vec{B}$ differ between these systems, the underlying physics we discuss here is the same.

\begin{figure*}[t]
{\includegraphics[width= 1\textwidth]{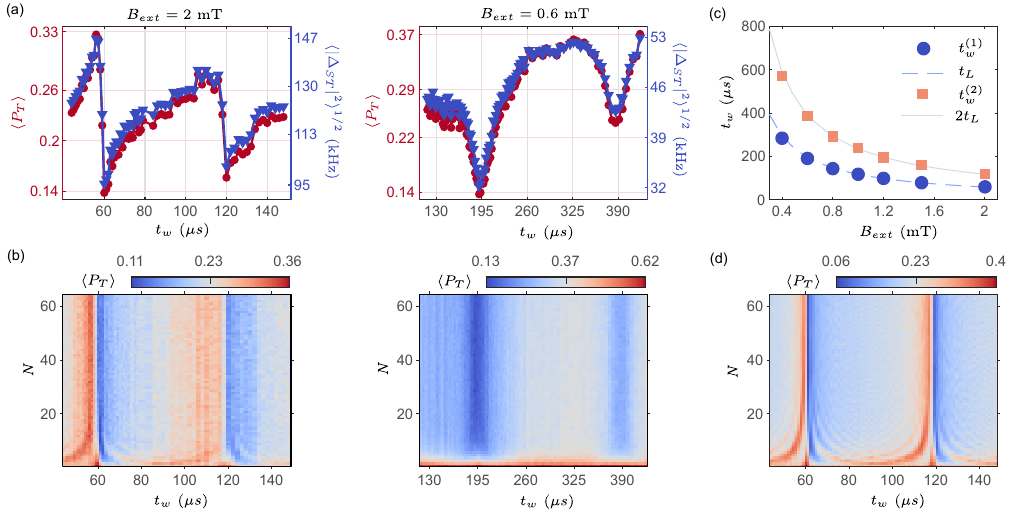}}
\caption{
\textbf{Screened dark state.}
(\textbf{a}) Triplet probability $\langle P_T \rangle$ (red) and root-mean-square gap  $\langle |\Delta|^2_{ST}\rangle^{1/2}$ (blue) vs. $t_w$ at $B_{ext}=2$ mT (left) and $0.6$ mT (right) after 128 LZ sweeps. The lines connecting the data points are guides to the eye. 
(\textbf{b}) $\langle P_T\rangle$  vs. LZ sweep number $N$ and $t_w$. After a few tens of LZ sweeps, the system reaches a steady state. (\textbf{c})  $t_w^{(1)}$, $t_w^{(2)}$, and $nt_L$ vs. $B_{ext}$ (\textbf{d}) Numerical simulations corresponding to the experiments in (b) at $B_{ext}=2$~mT with  $\Delta_{HF}=100$~kHz and $u=100$~kHz (see Supplementary Material).
}
\label{fig:Stroboscopic}
\end{figure*}
\section{Screened dark state}

Equation~\eqref{eq:ham} describes the evolution of an effective magnetic moment $\vec{S}$ interacting with another effective classical moment $\vec{v}$ as well as an effective magnetic field $\vec{B}$. Dynamic nuclear polarization can occur in this system via Landau-Zener (LZ) sweeps, where the detuning is swept from positive to negative. In the language of Eq.~\ref{eq:ham}, and when $u=0$, this corresponds to a change in the direction of $\vec{B}$ from $+z$ to $-z$. As this happens, $\vec{S}$ changes its orientation from $+z$, to the direction specified by $v_x$ and $v_y$, and then to $-z$. Near resonance, when $\vec{S}$ is oriented along transverse components of $\vec{v}$, the torque from $\vec{S}$ (the Knight field) rotates  $\vec{v}$ toward the $z$ axis. This motion of $\vec{v}$ is how dynamic nuclear polarization occurs in double quantum dots---the transverse polarization reduces in favor of the longitudinal polarization. After many LZ sweeps, the polarization vector $\vec{v}$ will gradually rotate out of the plane to point along the $z$ direction, and any transverse components will have vanished. This condition is the dark state we have previously reported~\cite{cai2025formation}, which features vanishing singlet-triplet coupling: $v_+\sim0$ [Fig.~\ref{fig:Stability} (e,f)]. 

These dynamics, however, can significantly change when $u>0$.  For example, previous research has shown that significant spin-orbit coupling can quench dynamic nuclear polarization~\cite{nichol2015quenching}. One can also see that the condition where $v_+ \sim~0$ is no longer a dark state, because the total singlet-triplet coupling $\Delta_{ST} = v_+ + u \neq 0$. However, a different kind of dark state can still occur if $v_+=-u$. In such a ``screened'' dark state, where the nuclear spins screen the spin-orbit coupling, electronic LZ sweeps cannot change the nuclear spin orientation because $\Delta_{ST} ~\sim 0$, and the nuclear spins are effectively phase locked  to the spin-orbit field.

An interesting complication to this intuitive picture results from the action of a real external magnetic field on the nuclear spins. (In this work, the field points in the out-of-plane direction.) Because the nuclear spins precess around this field according to their classical energy $-\vec{v}\cdot \vec{B}_{ext}$, this screening would occur once per Larmor period and will thus generally occur when the time between LZ sweeps $t_w$ is equal to the nuclear spin Larmor period. In the following, we verify that this prediction can indeed occur. We also show that by adjusting $t_w$, we can control the phase of the nuclear spin state and that we can use this state to measure the coherent precession of quantum-dot nuclear spins in different magnetic fields.

\begin{figure*}[t]
{\includegraphics[width= 1\textwidth]{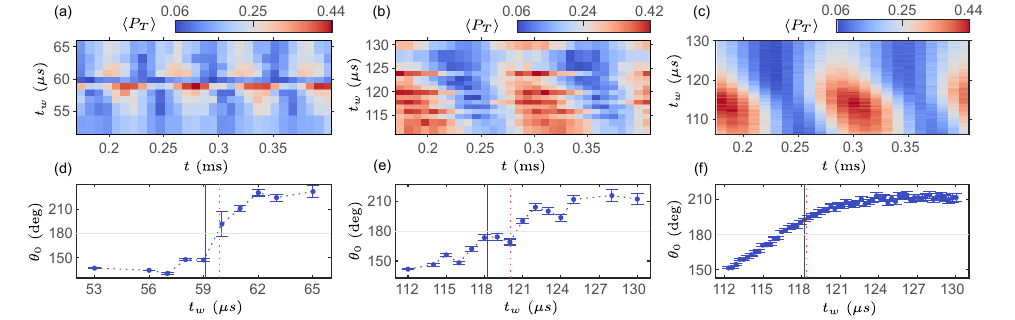}}
\caption{
\textbf{Nuclear-spin phase lock.}  $\langle P_T \rangle$ vs. $t_w$ and evolution time $t$ for
(\textbf{a}) $B_{ext}=2$ mT and (\textbf{b}) $B_{ext}=1$ mT. (\textbf{c}) Simulation at $B_{ext}=1$ mT and $u=50$ kHz, $\Delta_{HF} =100$ kHz. For the 2-mT data, there was a device shift at $t_w=60~\mu s$, which is responsible for the low visibility as well as slightly higher error bar at that $t_w$. Phase as a function of $t_w$ for the (\textbf{d}) 2-mT data, (\textbf{e}) 1-mT data, and (\textbf{f}) simulation, extracted by fitting to Eq.~\eqref{eq:pt}. In all of the phase plots, the horizontal light black line indicates $\theta_0=\pi$, while the black solid vertical line indicates $t_w=t_L$, where $\theta_0=\pi$ is expected. The dashed vertical line is the inverse of the mean fitted spin precession frequencies, with values of $120.0~\pm~1.5~\mu s$ and $59.9~\pm 0.2~\mu s$ for 1-mT and 2-mT data respectively. The slight discrepancy between the expected and fitted Larmor periods may be due to a field calibration error on the order of tens of $\mu$T. Error bars indicate the standard error from the fits.
}
\label{fig:phase}
\end{figure*}

To verify the nuclear spins can indeed screen the spin-orbit field, we vary the strength of the external magnetic field between 0.4~mT and 2~mT, and with a direction chosen such that $u$ has a similar magnitude to $v_+$ (see the Supplementary Material). Under these conditions, as we implement LZ sweeps, we expect the nuclear polarization vector to rotate up toward the $+z$ direction. As this happens however, the polarization vector should encounter the stable point with $v_+=-u$ and remain there [Fig.~\ref{fig:Stability}(g,h)]. Moreover, this stable configuration should occur when $t_w=t_L$.

We measure the LZ transition probability by initializing the ST qubit as a singlet, ramping $\epsilon$ through the avoided crossing, and then measuring $\langle P_T \rangle$, the average probability to measure the joint spin state after the sweep as a triplet. For fast LZ sweep rates, $\langle P_T \rangle\propto \langle|\Delta_{ST}|^2\rangle$, enabling us to extract the root-mean-square (rms) value of $\Delta_{ST}$~\cite{SHEVCHENKO2010Landau}. For reference, a typical LZ sweep changes the detuning approximately by 23 MHz in 30 $\mu$s. (In the language of Eq.~\eqref{eq:ham}, this ramp corresponds to changing the effective field $\vec{B}$ from pointing along the $+z$ direction to the $-z$ direction.) After the measurement, we include a variable wait to enable adjusting $t_w$~[Fig.~\ref{fig:Stability} (b)]. After 128 LZ sweeps, we include a longer wait, $t_{pause}$, of at least 2.8~ms to allow the nuclear spins to dephase and to add measurement calibration pulses~\cite{cai2025formation}. We also note that effects related to reduced singlet-triplet visibility in the presence of nuclear polarization~\cite{barthel2012relaxation} are not likely to occur in these experiments, because the dynamics explored here involve the polarized triplet state, instead of the unpolarized triplet, which can relax to the singlet during readout.

Figure~\ref{fig:Stroboscopic}(a) shows the results of these measurements at different magnetic fields. We observe a pronounced non-monotonic variation in $\langle P_T \rangle$ and $\langle |\Delta_{ST}|^2 \rangle ^{1/2}$ with $t_w$. In particular, $\langle P_T \rangle$ and $\langle |\Delta_{ST}|^2 \rangle ^{1/2}$ have local minima at equally-spaced times $t_w^{(1)}$, $t_{w}^{(2)}$, $\cdots$, which depend on $B_{ext}$. Plotting $t_w^{(1)}$ and $t_{w}^{(2)}$ vs. $B_{ext}$, together with the Larmor period $t_L$ and its integer multiples, reveals that $t_w^{(n)}\approx nt_L$ [Fig.~\ref{fig:Stroboscopic}(c)]. These data indicate that $\langle |\Delta_{ST}|^2 \rangle ^{1/2}$ reaches a local minimum through dynamic nuclear polarization when the wait between LZ sweeps is an integer multiple of the Larmor period. This configuration with a reduced $\langle |\Delta_{ST}|^2 \rangle ^{1/2}$ is a dark state because it features reduced electron-nuclear coupling. However, this state differs from our previous report~\cite{cai2025formation} in two respects. First, because $u\neq 0$, this dark state features an arrangement of nuclear spins such that the hyperfine coupling $v$ ``screens'' the spin orbit coupling $u$. Second, this screening is stroboscopic, because it occurs once per Larmor period, and only occurs when $t_w=nt_L$. In the next section below, we discuss data for the case that $t_w \neq t_L$.

Figure~\ref{fig:Stroboscopic}(b) shows the transient behavior of $\langle P_T \rangle$ as a function of LZ sweep number. At small sweep numbers, $\langle P_T \rangle$ oscillates before approaching its steady-state value. Figure~\ref{fig:Stroboscopic}(d) shows semiclassical numerical simulations with $B_{ext}=2$~mT and $\Delta_{HF}=100$~kHz, $u=100$~kHz, which agree well with the experimental observations. Here $\Delta_{HF}=\langle |v_+|^2\rangle^{1/2}$ characterizes the strength of the hyperfine coupling, where the average is over random nuclear spin states. We choose these parameters based on measurements of $\langle |\Delta_{ST}|^2 \rangle ^{1/2}$ as a function of magnetic field orientation (see the Supplementary Material for further details on these measurements and the simulations). These simulations calculate the quantum evolution of the electronic spins states, together with the classical evolution of the roughly 8000 nuclear spins in the system. The oscillations observed in both experiment and simulation result from the vector $\vec{v}$ spiraling as it rotates up toward the $+z$ direction and eventually encounters the dark state.

\begin{figure*}[t]
{\includegraphics[width= 1\textwidth]{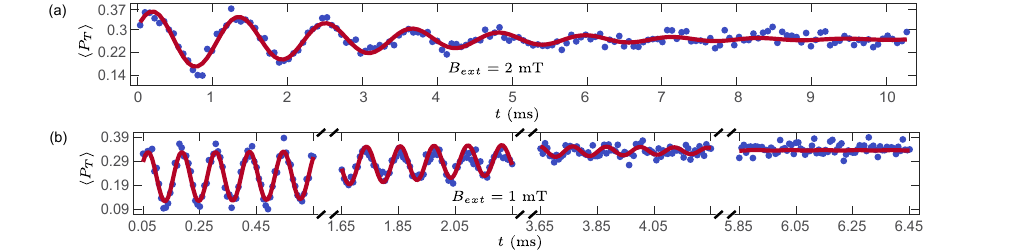}}
\caption{
\textbf{Nuclear-spin coherence time.} 
(\textbf{a}) $\langle P_T\rangle$ vs. $t$ at $B_{ext}=2$ mT and fit to Eq.~\eqref{eq:pt}. At $t=0$, the spin ensemble is prepared with $\theta_0=\pi$ with $t_w=t_L$. The oscillations are aliased, because we use a time step of $\Delta t=64~\mu$s to acquire the data. The fit yields $T_2^*=3.7 \pm 0.3$~ms and $\gamma=1.4 \pm 0.2$ (\textbf{b}) $\langle P_T\rangle$ vs. $t$ at $B_{ext}$=1~mT and fit to Eq.~\eqref{eq:pt}. $T_2^*=3.0 \pm 0.1$~ms and $\gamma=2.1 \pm 0.2$. Here there is no aliasing ($\Delta t=8\mu s$). Uncertainties are standard errors associated with the fits.
}
\label{fig:coherence}
\end{figure*}

\section{Nuclear-spin phase lock}
Having established that the nuclear spins can stroboscopically screen the spin orbit coupling, we now demonstrate the ability to phase-lock the nuclear spins. In particular, we show that this pumping process results in a definite, controllable phase for the nuclear spin ensemble, and that this phase can be controlled. Moreover, this phase-locking occurs without any microwaves or phase references and makes use only of the internal spin-orbit coupling as a reference. 

As an initial example, we demonstrate the ability to measure the free evolution of the nuclear spins. For this we use a ``pump-evolve-probe'' sequence~[Fig.~\ref{fig:Stability}(c)]. After pumping the nuclear spins into the screened dark state, we pause for a variable length of time $t$, and then we conduct a final ``probe'' LZ sweep to measure the $t$-dependent triplet return probability. Additional wait pulses after the probe pulse ensure that total repetition rate, $t_{tot}$, for the experiment is the same for different pause times. 

To understand the expected result of such an experiment, suppose that at $t=0$ the nuclear spin ensembles of dots 1 and 2 have magnitudes $v_1$ and $v_2$ and phases $\theta_1$ and $\theta_2$ with respect to $u$. Then, $\Delta_{ST}(t)=v_1e^{i(\omega_1t+\theta_1)}-v_2e^{i(\omega_2t+\theta_2)}+u$ where $\omega_1$ and $\omega_2$ denote the nuclear-spin precession frequencies. (In general, these will both equal the Larmor frequency up to inhomogeneous dephasing effects.) We find that (see the Supplementary Material), 
\begin{equation}
    \langle P_T \rangle=A_0+A_1e^{-2(t/T_2^*)^\gamma}+A_re^{-(t/T_2^*)^\gamma}\cos(\omega_L t+\theta_0).
    \label{eq:pt}
\end{equation}
Here $\theta_0$ is an initial phase that depends on $\theta_1$ and $\theta_2$. $T_2^*$ is the inhomogeneous nuclear spin dephasing time, and $\gamma$ is an exponent that depends on the nuclear spin decay envelope. $A_0$, $A_1$, and $A_r$ are constants that depend on the initial nuclear spin configuration. $\omega_L$ is the Larmor frequency, which we assume is the same for both dots.

In Fig.~\ref{fig:phase}(a) we present results of such a  measurement performed at $B_{ext}=2$ mT. As expected the LZ transition probability of the last probe LZ sweep varies sinusoidally in time, as the nuclear spins undergo free precession. Moreover, the observed  frequency matches the expected Larmor frequency of the nuclear spins. In previous experiments with quantum-dot nuclear spins, the Larmor precession frequency of the nuclear spins could be observed through correlation-based measurements~\cite{nichol2015quenching}. Here, however, we observe the Larmor precession without the need to correlate the spin polarization at different times, because we can prepare the spin ensemble with a definite phase. 

Interestingly, we find that $\theta_0$ depends on $t_w$ [Fig.~\ref{fig:phase}(d)]. When $t_w=t_L$, $\theta_0 \approx \pi$, as it must be to sustain the dark state. However, when  $t_w < t_L$ we find $\theta_0 < \pi$, and when $t_w > t_L$, we observe $\theta_0 > \pi$. We observe similar behavior at $B_{ext}=1$ mT, as shown in Figs.~\ref{fig:phase}(b,e), and our simulations show similar behavior for both the nuclear-spin free evolution and the oscillation phase as shown in Figs.~\ref{fig:phase}(c,f). The modest discrepancy between simulation and experiment likely reflects differences in the spin–orbit coupling strength and electrical noise parameters used in the model (see Supplementary Material).

An intuitive reason for the dependence of $\theta_0$ on $t_w$ is that the total phase acquired by the nuclear spin polarization as a result of the LZ sweep, together with the free evolution, must equal 2$\pi$ for each LZ sweep and wait, to reach a steady state. When $\theta_0=\pi$, the rotation induced by the Knight field vanishes, because $\Delta_{ST} \sim 0$ and the electrons remain in the singlet state. Thus the net phase  from Larmor evolution must equal 2$\pi$, which occurs when $t_w=t_L$. When $\theta_0 \neq \pi$, the Knight field will induce a non-zero rotation. In this case, the stable state requires that the phase acquired during Larmor evolution differs from 2$\pi$, and thus that $t_w \neq t_L$.

\section{Nuclear-spin coherence times}

We now show that the nuclear-spin phase lock can be used to measure the nuclear-spin coherence time and envelope. We use the same ``pump-evolve-probe'' sequence, except that now the ``evolve'' time ranges from $0-10$ ms.  Figure \ref{fig:coherence}(a) shows the result of this measurement for $B_{ext}=2$ mT, 
and we resolve the entire coherence envelope of the nuclear-spin ensemble. By fitting the data to Eq.~\eqref{eq:pt}, we find the dephasing time to be $3.69\pm0.32$~ms with a decay exponent of $\gamma=1.4\pm0.2$, in agreement with previously reported values for quantum-dot nuclear spin coherence times~\cite{pla2014coherent,cai2025formation}. We emphasize that this measurement represents the coherence of the nuclear spins interacting with the electrons in the double quantum dot, and is thus a direct probe of the magnetic environment of the electrons in the quantum dots. We also emphasize that the net polarization of the ensemble  is on the order of $\sqrt{10^3}-\sqrt{10^4}=30-100$ nuclear spins. Intriguingly, we find a lower coherence time of $3.03\pm0.09$~ms and different decay exponent $\gamma=2.1 \pm 0.2$ at $B_{ext}=1$~mT. The origin of this difference will be the subject of future work and may have implications for electron-spin qubits that operate in low fields, such as exchange-only qubits~\cite{DiVincenzo2000Universal}.    

\section{Discussion and outlook}
We have presented evidence for a nuclear-spin dark state that forms when the nuclear polarization cancels the electronic spin orbit coupling in a silicon singlet-triplet qubit. Because the nuclear spin phase is set by the electronic spin orbit coupling, this process results in a deterministic, controllable phase for the nuclear spins without the use of any microwaves or external references, and which can be used to enable measuring their coherence. 

Not only does this work verify long-standing predictions about collective nuclear spin dynamics but it also expands the range of possible tools for controlling quantum-dot nuclear spin ensembles.  Previous theoretical work, for example, has suggested the possibility of a topological phase transitions~\cite{Rudner2009Topological,Rudner2010Phase}, increased efficiency of dynamic nuclear polarization~\cite{Brataas2011Nuclear}, and exotic interference effects~\cite{Rudner2010Phase-sensitive} when both spin-orbit coupling and hyperfine interactions compete. Our work is a first step in exploring these tantalizing prospects in solid-state nuclear spin systems. On a broader level, our work explores how both internal and external control fields can modify how dynamic nuclear polarization occurs, shedding new light on this essential tool for quantum information science, solid-state physics and chemistry, and medicine.  

\section{Acknowledgments}
We thank L. F. Edge of HRL Laboratories for the epitaxial growth
of the SiGe material and E. J. Connors for device fabrication. This work was sponsored by the Army Research Office under grant No. W911NF-23-1-0115. The views and conclusions contained in this document are those of the authors and should not be interpreted as representing the official policies, either expressed or implied, of the Army Research Office or the U.S. Government. The U.S. Government is authorized to reproduce and distribute reprints for Government purposes notwithstanding any copyright notation herein.

\end{document}


\title{Supplementary Material for \\ ``Phase locking nuclear spins in silicon with spin-orbit coupling"}

\author{Habitamu Y. Walelign}
 \thanks{These authors contributed equally.}

 \author{Manas Ranjan Sahu}
 \thanks{These authors contributed equally.}

 \affiliation{Department of Physics and Astronomy, University of Rochester, Rochester, NY 14627, USA}
 \affiliation{University of Rochester Center for Coherence and Quantum Science, Rochester, NY 14627, USA}

 \author{John M. Nichol}
 \email{john.nichol@rochester.edu}
 \affiliation{Department of Physics and Astronomy, University of Rochester, Rochester, NY 14627, USA}
\affiliation{University of Rochester Center for Coherence and Quantum Science, Rochester, NY 14627, USA}
\maketitle

\tableofcontents
\clearpage
\section{Methods} 
\subsection{Spin orbit coupling characterization}
\begin{figure}[ht]
    \centering
    \includegraphics[width=1\linewidth]{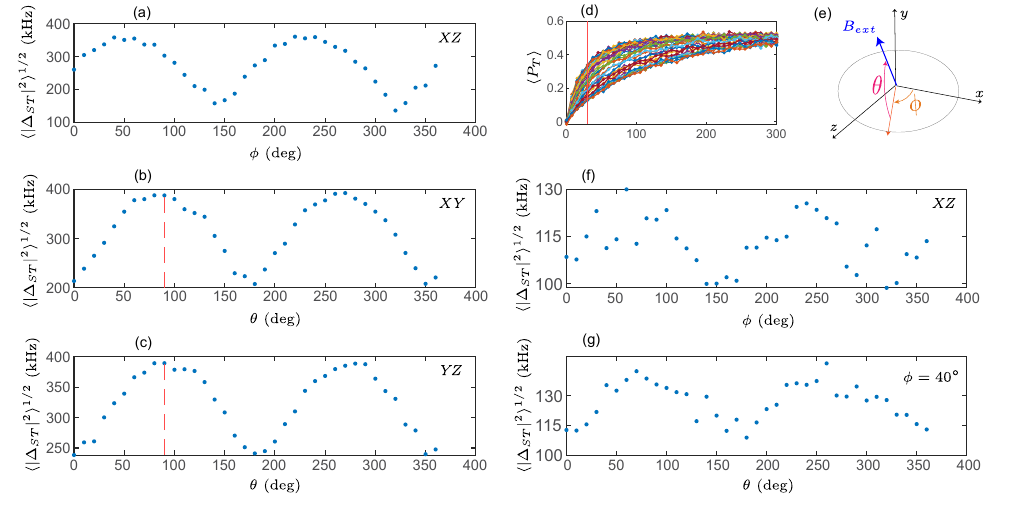}
    \caption{\textbf{SOC characterization}:
    RMS magnitude of $\Delta_{ST}$ vs. external field direction at 10 mT in the (\textbf{a}) $XZ$ plane, (\textbf{b}) $XY$ plane and (\textbf{c}) $YZ$ plane. The $XZ$-plane data were taken at a different time from the other data sets, and the difference in $\langle \Delta_{ST}^2 \rangle^{1/2}$ along the $x$-axis ($\phi=0$ and $\theta=\pi/2$) likely results from hysteresis in our magnet.
    (\textbf{d}) LZ transition probability at different sweep times for the $YZ$-plane data. The different colors correspond to the different angles. The time used in the analysis for (a,b,c) is shown by the vertical red line. (\textbf{e}) Coordinate system used here. The axis connecting the two quantum dots is parallel to the $z$-axis, and the substrate lies in the $XZ$ plane. RMS value of $\Delta_{ST}$ vs. external field direction at 2~mT in the (\textbf{f}) XZ plane and (\textbf{g}) at $\phi=40^{\circ}$ sweeping $\theta$.}
    \label{figS:socChar}
\end{figure}

We characterize $u$ and $\Delta_{HF}$ in our device using the procedure described in Ref.~\cite{nichol2015quenching}. In brief, we perform LZ sweeps through the singlet-triplet resonance point at different orientations of the external magnetic field. For fast sweeps, the LZ transition probability is proportional to $\langle|\Delta_{ST}|^2\rangle^{1/2}$. Figure \ref{figS:socChar} shows the results of these measurements at $|B_{ext}|=10$~mT and $|B_{ext}|=2$~mT. We expect that $\langle|\Delta_{ST}|^2\rangle^{1/2}$ oscillates with magnetic field, reaching a maximum value when the external field is perpendicular to the internal spin-orbit field, and when $u$ is largest. The minimum value of $\langle|\Delta_{ST}|^2\rangle^{1/2}$ occurs when the spin-orbit field is parallel to the external field and $u$ is minimum. We found that the out-of-plane axis, which we define as the $y$-axis, corresponds to the direction with the largest $u$.  This direction is used in all of the experiments in this work. 

To get the magnitudes of $\Delta_{HF}$ and $u$ used in the main text, we used the 2-mT measurement shown in Fig.~\ref{figS:socChar}(f,g). We extract the interdot tunnel coupling $t_c$ as $t_c=J(0)$ where $J(\epsilon)$ is obtained by fitting a spin-funnel measurement that maps $J(\epsilon_{ST})=\bar{g}\mu_B B_{ext}$~\cite{cai2023coherent}. Using the extracted tunnel coupling $t_c=2.36$~GHz, we calculate the singlet mixing angle as $\theta_{mix}=\tan^{-1}(B_{ext}/t_c)$, which quantifies the $\ket{(2,0)S}$ and $\ket{(1,1)S}$ mixture as  $\ket{S}=\cos{(\theta_{mix})}\ket{(1,1)S}+\sin{(\theta_{mix})}\ket{(2,0)S}$. Then we calculated the magnitudes of $u$ and $\Delta_{HF}$ using $\langle|\Delta_{ST}|^2\rangle^{1/2}=\delta B_\perp\cos{(\theta_{mix})}+ \Omega_{so}\sin{(\theta_{mix})}\sin{(\phi-\phi_0)}$ where $\phi$ is the in-plane field direction with $\phi_0$ describing an offset, and $\Delta_{HF}=\delta B_{\perp}\cos(\theta_{mix})$ and $u=\Omega_{so}\sin(\theta_{mix})$. Here $\delta B_\perp$ describes the root-mean-square (rms) transverse hyperfine field difference between the two dots and $\Omega_{so}$ is the spin-orbit matrix element between the (2,0) singlet and the triplet. From the 2-mT data we estimate the rms magnitude of $\Delta_{ST}$ and $\Delta_{HF}$ to be about 142~kHz and 100~kHz respectively. From this we compute $\delta B_\perp$ and $\Omega_{so}$ and assume that these do not change between 2~mT and 1~mT since these describe the rms magnitudes of the hyperfine gradient and maximum spin orbit matrix element before any projection. By calculating the mixing angle at 1~mT, we compute the corresponding $u$ and $\Delta_{HF}$. We note that the 2-mT data in Fig.~\ref{figS:socChar} was taken at a second cool down and slightly different tuning than the data in the main text, while the 10-mT data was taken during the same cool down. Extracting $\delta B_\perp$ and $\Omega_{so}$ from the 10-mT data yielded values larger than those from the 2-mT data, which could be due to an internal spin-orbit field  with an out-of-plane component~\cite{zhang2020giant}. Our expression for $\langle |\Delta_{ST}^2\rangle^{1/2}$ as function of $\phi$ assumes the spin-orbit field lies in the plane. Thus, we expect that the 2-mT data, with a smaller mixing angle, is less susceptible to any out-of-plane component and yields more accurate values.

\subsection{Phase determination}
\label{sec-Phase_determination}
Here we derive Eq.~(3) in the main text. Let $v_{1(2)}$ and $\theta_{1(2)}$ denote the magnitudes and phases of the transverse polarizations in dots 1 and 2 at $t=0$. Then,
\begin{equation}
    \Delta_{ST} (t)=v_1e^{i(\omega_1t+\theta_1)}-v_2e^{i(\omega_2t+\theta_2)}+u
    \label{Eq.S1}
\end{equation}
where $\omega_{1(2)}$ is the nuclear-spin precession in dot 1(2). We account for nuclear spin dephasing by averaging Eq.~\eqref{Eq.S1} over Gaussian distributions of precession frequencies for each dot, with means  $\bar{\omega}_{1(2)}$ and mean standard deviations $\sigma_{1(2)}$. Then, we have
\begin{equation}
    \begin{split}
        \langle |\Delta_{ST}(t)|^2\rangle = & \int_{-\infty}^{\infty} P(\omega)|\Delta_{ST}(t)|^2d\omega\\
        =&v_1^2+v_2^2+u^2 \\
        &-\frac{2v_1v_2}{2\pi\sigma_1\sigma_2}\Re{\int_{-\infty}^{\infty}e^{i[(\omega_1-\omega_2)t+\theta_1-\theta_2]}e^{-\frac{(\omega_1-\bar{\omega}_1)^2}{2\sigma_1^2}-\frac{(\omega_2-\bar{\omega}_2)^2}{2\sigma_2^2}}d\omega_1d\omega_2}\\
        &+\frac{2u}{2\pi\sigma_1\sigma_2}\Re{v_1\int_{-\infty}^{\infty}e^{i(\omega_1 t+\theta_1)}e^{-\frac{(\omega_1-\bar{\omega}_1)^2}{2\sigma_1^2}}d\omega_1-v_2\int_{-\infty}^{\infty}e^{i(\omega_2 t+\theta_2)}e^{-\frac{(\omega_2-\bar{\omega}_2)^2}{2\sigma_2^2}}d\omega_2}.
    \end{split}
    \label{Eq.S3}
\end{equation}
Assuming independent fluctuations between dots, 
\begin{equation}
    \begin{split}
        \langle |\Delta_{ST}(t)|^2\rangle=&v_1^2+v_2^2+u^2 \\
        -& 2v_1v_2e^{\frac{-(\sigma_1^2+\sigma_2^2)t^2}{2}}\cos[(\bar{\omega}_1-\bar{\omega}_2)t+\theta_1-\theta_2]\\
        +& 2u[v_1e^{-\frac{\sigma_1^2t^2}{2}}\cos(\bar{\omega}_1t+\theta_1)-v_2e^{-\frac{\sigma_2^2t^2}{2}}\cos(\bar{\omega}_2t+\theta_2)].
    \end{split}
    \label{Eq.S4}
\end{equation}
Setting $\bar{\omega}_1=\bar{\omega}_2\equiv\omega$ and $\sigma_1=\sigma_2\equiv\sigma$, the above equation can be rewritten as,
\begin{equation}
    \begin{split}
        \langle|\Delta_{ST}(t)|^2\rangle = & A_0+A_1e^{-2\left(\frac{t}{T_2^*}\right)^2}+[A_c\cos(\omega t)+A_s\sin(\omega t)]e^{-\left(\frac{t}{T_2^*}\right)^2}\\
        =&A_0+A_1e^{-2\left(\frac{t}{T_2^*}\right)^2}+A_re^{-\left(\frac{t}{T_2^*}\right)^2}\cos(\omega t+\theta_0),
    \end{split}
    \label{Eq:fit}
\end{equation}
where
\begin{align}
    A_0 &\equiv v_1^2+v_2^2+u^2 \label{Eq.A0}\\
    A_1 & \equiv -2v_1v_2\cos(\theta_1-\theta_2) \label{Eq.A2}\\
    A_c & \equiv 2u[v_1\cos(\theta_1)-v_2\cos(\theta_2)]\label{Eq.Ac}
    \\
    A_s &\equiv -2u[v_1\sin(\theta_1)-v_2\sin(\theta_2)] \label{Eq.As}\\
    \theta_0&= \tan^{-1}\left(-\frac{A_s}{A_c}\right)\label{Eq.theta0} \\
    A_r &=\sqrt{A_c^2+A_s^2} \label{Eq.Ar}
\end{align}
with $T_2^*\equiv \frac{\sqrt{2}}{\sigma}$. 
Equation \ref{Eq:fit} is related to the fitting expression used in determining the phase in the main text. The exact expression in the main text is obtained if one assumes a characteristic function for the noise average of the form $\mathbb{E}[e^{iw_it}]=e^{i\bar{\omega_i}t-\left(\frac{t}{T_{2,i}^*}\right)^{\gamma_i}}$,  where $\mathbb{E}[{e^{i\omega_it}}]\equiv\int_{-\infty}^\infty P(\omega)e^{i\omega_it}d\omega$ defines the expectation value of $e^{i\omega_it}$ with respect to the probability distribution $P(\omega)$. This reduces to the 
Gausssian case when $\gamma=2$. When $t\ll T_2^*$, Eq.~\ref{Eq:fit} reduces to a sinusoidal oscillation with an offset as expected.

\clearpage
\begin{figure}
    \centering
    \includegraphics[width=1\linewidth]{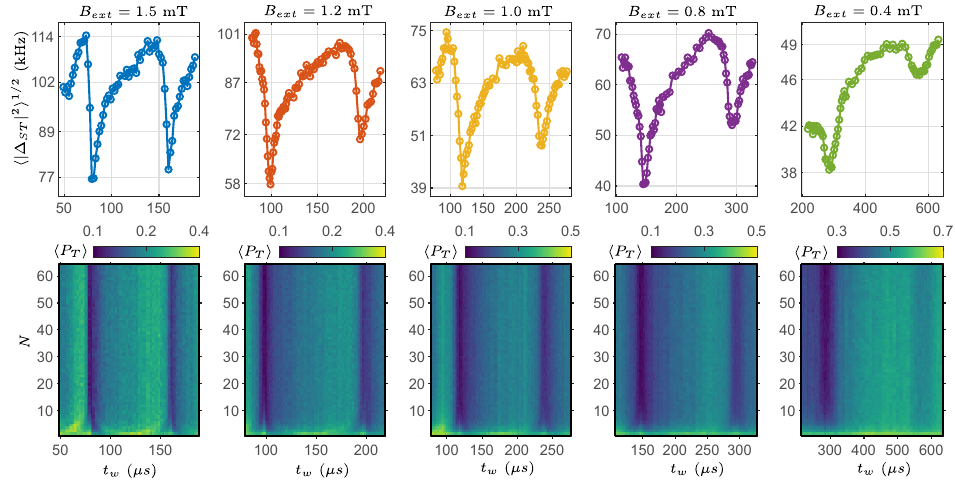}
    \caption{
    \textbf{Additional data for screened dark state}: 
    (\textbf{top row}) Steady-state values of $\langle |\Delta_{ST}|^2\rangle^{1/2}$, as a function of the time between successive LZ sweeps for different fields. 
    (\textbf{bottom row}) LZ transition probability vs sweep number and wait time.  At low fields, the steady state occurs rapidly, while at higher fields, oscillatory behavior emerges.}
    \label{figS:Stroboscopic}
\end{figure}

\clearpage
\section{Simulation}
\subsection{Hamiltonian}
In this section, we review the theoretical model used in our simulation. Our model is similar to that described in previous work~\cite{cai2025formation}, except that here we include the spin-orbit matrix element $u$. In brief, we integrate the time-dependent Schr\"odinger equation for the electron-nuclear system with a quantum mechanical description of electrons and a semiclassical approximation for the nuclear spins, where we account for both the external magnetic field and the Knight field. Within the semiclassical approximation and for operations near $\epsilon_{ST}$, the Hamiltonian for the effective two-electron system in the reduced subspace spanned by  $\ket{T_-}$ and $\ket{S}$ can be captured by
\begin{equation}
    H=\begin{pmatrix}
    \Delta E_{ST_-}+v_z(t) & u+v_+(t)\\
    u+v_-(t) & 0
    \end{pmatrix},
    \label{S1}
\end{equation}
where the matrix elements above are defined as
\begin{align}
    v_z &\equiv\bra{T_-}H_{hf}\ket{T_-}\label{S2}\\
    v_+&\equiv\bra{T_-}H_{hf}\ket{S}=v_-^\dagger\label{S3}\\
    u&\equiv\bra{T_-}H_{SO}\ket{S}.
    \label{S4}
\end{align}
The hyperfine Hamiltonain is 
\begin{align}
    H_{hf}=\bar A\sum_{k=1}^N\sum_{i=1}^2\delta(\textbf{R}_k-\textbf{r}_i)\textbf{I}_k\cdot\textbf{S}_i,
\end{align}
where $\textbf{R}_k$ specifies the nuclear spin coordinates for each of the nuclear spins and $\textbf{r}_i$ specifies the coordinate for the two electrons. $\bar A$ is the effective electron-nuclear hyperfine coupling constant and we used $\bar A=\frac{4\Delta_{HF}}{I\sqrt{\frac{1}{N_1}+\frac{1}{N_2}}}$ in this work where $N_1$ and $N_2$ denote the number of nuclear spins in dots 1 and 2 respectively and I denotes the spin magnitude. $H_{SO}$ is the spin orbit coupling Hamiltonian that in general has both Rashba and Dresselhaus contributions. $\Delta E_{ST_-}=J-\bar g\mu_BB^z$ is the $\ket{S}-\ket{T_-}$ detuning in the absence of hyperfine effects. $J$ denotes the exchange coupling [see Fig. \ref{fig:Stability} (d)]. We approximate the effect of the electron spins on the nuclei through the classical approximation of the Overhauser field ~\cite{Neder2014Theory}. Thus, the time dependence of $v_+$ results  from the influence of the external magnetic field and the Overhauser field on the nuclear spins.

Using a simple capacitance model for flat disks and the measured lever arms of the two plunger gates defining the dots, we estimate the radial size of the dots to be 29 and 38~nm, from which we estimate the number of nuclear spins to be around 8000. To account for nuclear-spin dephasing, we add effective white magnetic noise to the external field in our simulations, resulting in an inhomoegeneously broadened coherence time of 4~ms.  We simulate nuclear spin relaxation by using a random re-initialization of the nuclear spins with a probability per unit time of 0.2~s$^{-1}$, corresponding to a relaxation time that we chose to be 5 seconds~\cite{cai2025formation}. Finally, a quasi-static and Gaussian distributed electrical noise with an rms value of 460 kHz is added in the detunings to account for voltage fluctuations in the experiment. Each of the simulations have been averaged over 500 runs with nuclear spin relaxation captured after a full pump-pause sequence.

\subsection{Effects of $u$ and $\Delta_{HF}$ on steady state behavior}
\begin{figure}
    \centering
    \includegraphics[width=1\linewidth]{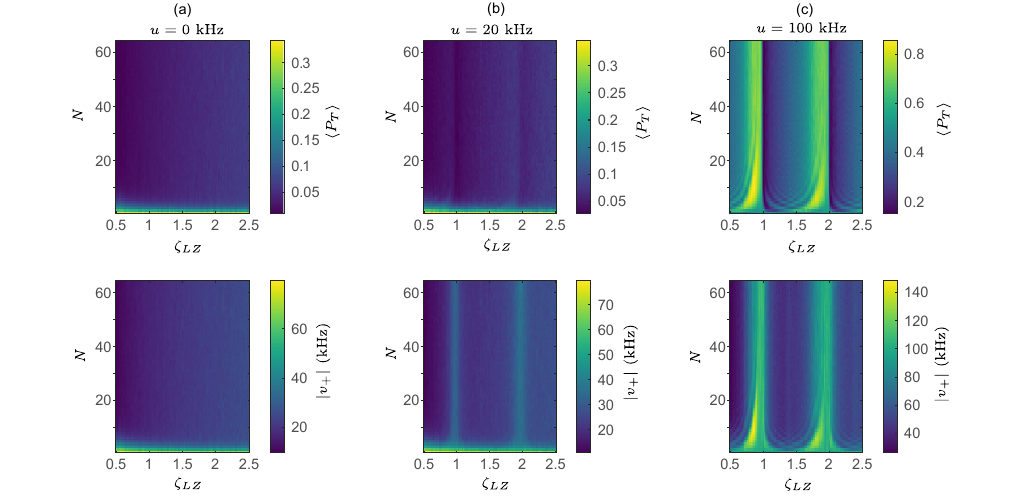}
    \caption{\textbf{Spin orbit coupling effect}: 
    The presence of the stroboscopic effect depends on the magnitude of the spin orbit coupling. The magnitude of $v_+$ will adjust (at steady state) to a value comparable to the spin orbit coupling so that at resonance it screens the spin orbit coupling. Here $\zeta_{LZ}=t_{w}/t_{L}$. The top row shows $\langle P_T \rangle$ while the bottom shows $v_+$. These simulations are done for $B_{ext}=1$~mT and $\Delta_{HF}=100$~kHz. The columns are results for 
    (\textbf{a}) $u=0$, (\textbf{b}) $u=20$~kHz, and (\textbf{c}) $u=100$~kHz.}
    \label{figS:Chevron}
\end{figure}

To understand how the system approaches the screened dark state using our numerical model, we fix $\Delta_{HF}=100$ kHz and simulated the effects of changing $u$ on the system dynamics. Figure \ref{figS:Chevron}(a) shows the formation of the conventional dark state when $u=0$~\cite{cai2025formation}.  Figure \ref{figS:Chevron}(b) shows the case with $u=20$ kHz. Here, the system takes longer to reach a steady state, and $|v_+|$ approaches $u$ on resonance. Figure~\ref{figS:Chevron}(c) shows the case where $u=100$ kHz. Now, on resonance $|v_+|$ does not reduce but instead maintains its value to cancel $u$. Near resonance, oscillatory behavior emerges.

\begin{figure}
    \centering
    \includegraphics[width=1\linewidth]{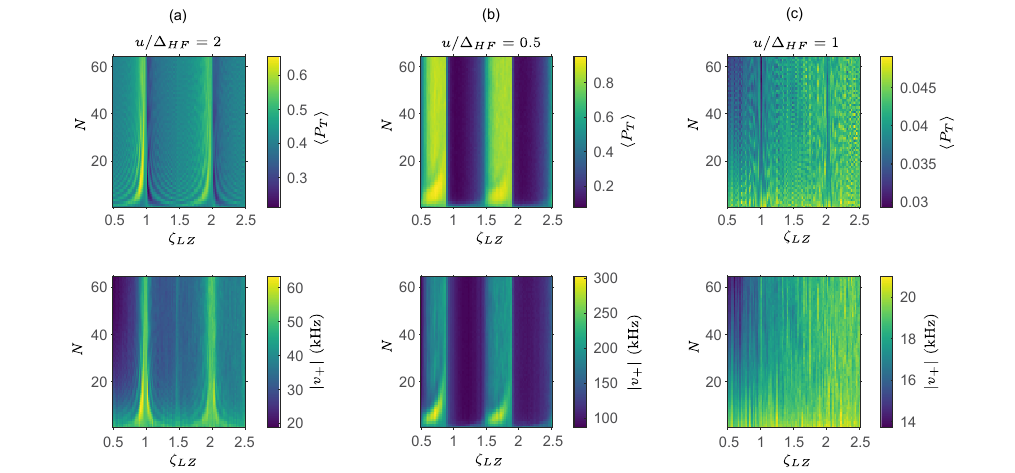}
    \caption{\textbf{Competition between SOC and HF interaction}: Demonstration of the competition between the spin orbit coupling (SOC) and hyperfine interaction (HF) that leads to the oscillatory behavior before steady state is achieved. These simulations are performed with $B_{ext}=1$~mT. The first row shows how $\langle P_T\rangle$ oscillates before achieving steady state. The second row shows how  $v_+$ evolves. In column (\textbf{a}) $u=100$~kHz, $\Delta_{HF}=50$~kHz. On resonance, $|v_+|$ maintains the maximum possible value to reduce $\Delta_{ST}$ as much as possible. In column (\textbf{b}), $u=100$ kHz and $\Delta_{HF}=200$ kHz. On resonance, $|v_+|$ reduces to screen $u$. In (\textbf{c}) $u=\Delta_{HF}=$ 20 kHz. On resonance, $|v_+|$ maintains its initial value to screen $u$. Comparing the three columns, we see with increasing $u$ relative to $\Delta_{HF}$, oscillatory behavior emerges.
    }
    \label{figS:competition}
\end{figure}

We also simulated the effects of changing $\Delta_{HF}$ relative to $u$. Figure \ref{figS:competition}(b) shows the ``screened state'' where $|v_+|$ saturates to $u$ at resonance instead of continuously reducing its magnitude below $u$. We invite particular attention to Fig.~\ref{figS:competition}(a) where $\Delta_{HF}=u/2$. In this case, $v_+$ oscillates, because it cannot screen $u$. When both $\Delta_{HF}=u=20$ kHz, the transition probability is very small as expected but the robust stroboscopic screening is still visible. Once again we see oscillations in this case as shown in Fig.~\ref{figS:competition}(c). All simulations shown in this work show how robust this resonance condition is and reproduced the experimentally observed oscillations and their important characteristics.

\subsection{Adiabaticity and steady state gap}
\begin{figure}
    \centering
    \includegraphics[width=1\linewidth]{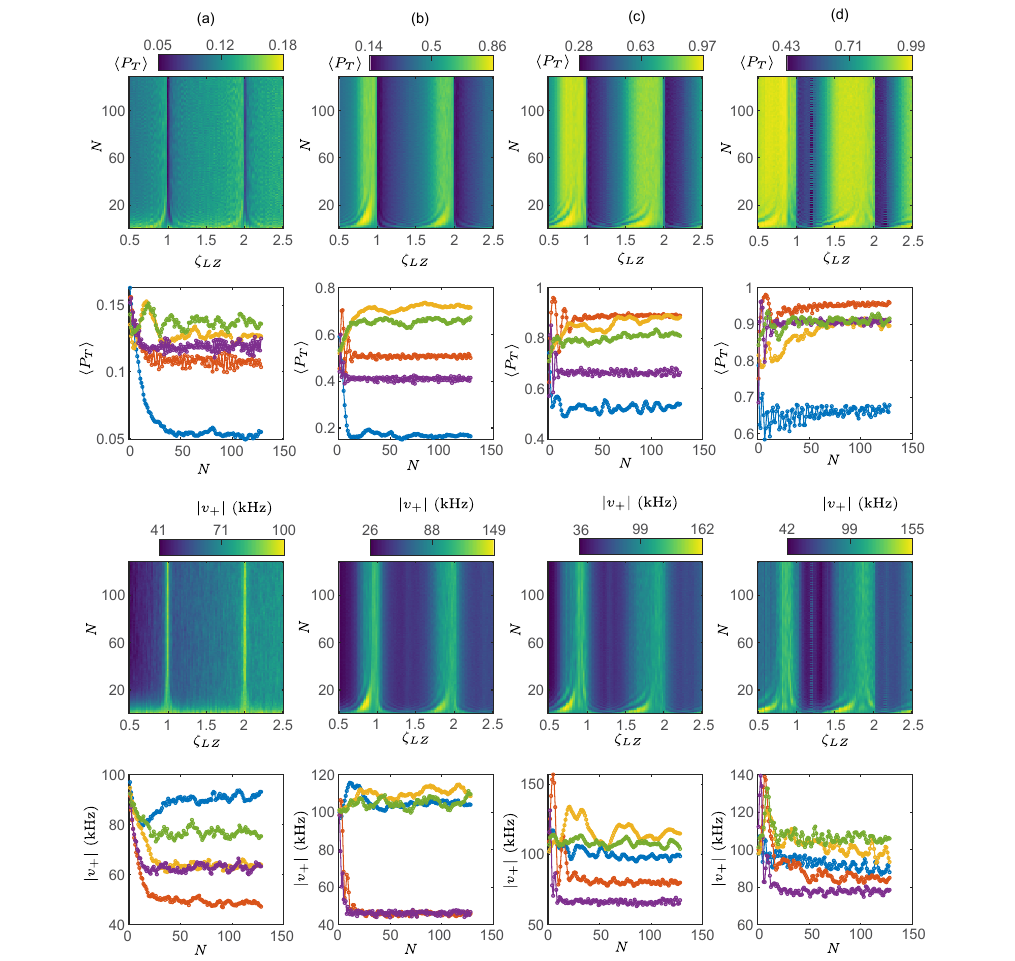}
    \caption{\textbf{Effect of LZ sweep rate on nuclear dynamics}: All of the simulations here have the same parameters except the LZ sweep rate $\propto \Delta\epsilon/t_{LZ}$ where $\Delta\epsilon$ is the detuning change defining the LZ sweep window and $t_{LZ}$ is the sweep time for this detuning in this window. Shown in each column are results for different sweep rates with \textbf{(a)} $t_{LZ}=5~\mu$s, \textbf{(b)} $t_{LZ}=30~\mu$s, \textbf{(c)} $t_{LZ}=60~\mu$s and \textbf{(d)} $t_{LZ}=100~\mu$s. The plots in the second and last rows are selected data from the 2D plots in the first and third rows, respectively, to show the variation in the behavior with sweep rate. $\zeta_{LZ}=t_{w}/t_{L}=1$ (blue), 0.7 (red), 0.96 (orange), 1.5, (purple), and 1.96 (green).
    }
    \label{figS:Rate}
\end{figure}

To explore how the results we present depend on the LZ sweep rate, we performed numerical simulations with different LZ sweep times. Figure \ref{figS:Rate} shows simulations with different LZ sweep rates for $u=\Delta_{HF}=100$~kHz, $t_{pause}=5$~ms, $B_{ext}=1$~mT and all other parameters kept the same. In all cases, we observe a robust stroboscopic effect. However, the precise behavior away from the resonance condition depends on the sweep rate. 

\subsection{Effect of $u$ and $\Delta_{HF}$ on the phase}
\begin{figure}
    \centering
    \includegraphics[width=1\linewidth]{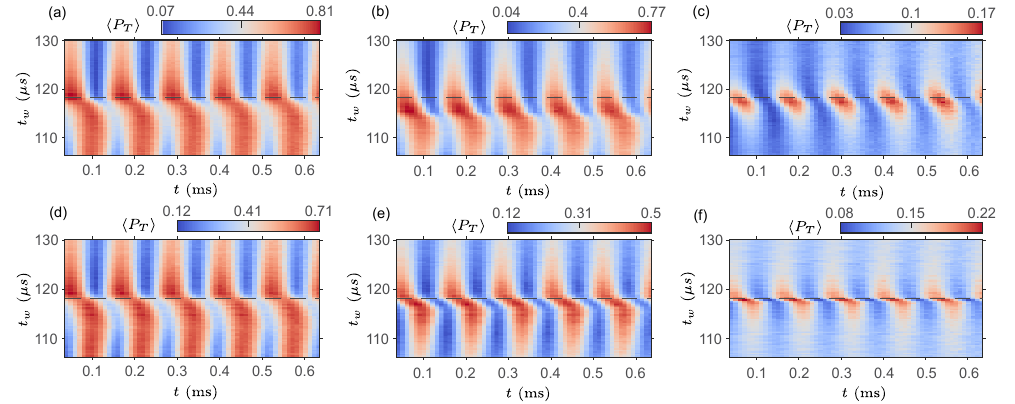}
    \caption{\textbf{Effect of magnitude of $u$ and $\Delta_{HF}$ on the nuclear spin phase}: Simulations for different values of $u$ and $\Delta_{HF}$ with all other parameters kept the same. Shown are for (\textbf{a}) $u=100$ kHz, $\Delta_{HF}=100$ kHz (\textbf{b}) $u=75$ kHz, $\Delta_{HF}=100$ kHz (\textbf{c}) $u=25$ kHz, $\Delta_{HF}=50$ kHz (\textbf{d}) $u=100$ kHz, $\Delta_{HF}=75$ kHz (\textbf{e}) $u=75$ kHz, $\Delta_{HF}=50$ kHz and (\textbf{f}) $u=50$ kHz, $\Delta_{HF}=25$ kHz. In all plots, the black dashed line shows the $t_w$ value where the stroboscopic condition is met and the phase is $\theta_0=\pi$.  
    }
    \label{fig:spinOrbitonPhase}
\end{figure}

While $u$ does not appear explicitly in  the expression for $\theta_0$ in Eq.~\ref{Eq.theta0}, the initial values $v_1$ and $v_2$, which are the steady state transverse polarizations during pumping, will depend on $u$ and $\Delta_{HF}$. To illustrate this dependence, Fig.~\ref{fig:spinOrbitonPhase} shows simulated pump-pause-probe experiments for different values of $u$ and $\Delta_{HF}$. In general, we find that when $u > \Delta_{HF}$, $\theta_0$ changes more rapidly.  This is consistent with the notion that $u$ determines the phase of the nuclear spin ensemble. 
